\documentclass[preprint2]{aastex6}

\usepackage{booktabs}
\usepackage{longtable}
\usepackage{xspace}
\usepackage[T1]{fontenc}
\usepackage{listings}
\usepackage{amsmath}

\definecolor{lbcolor}{rgb}{0.9,0.9,0.9}
\newcommand\latex{La\TeX}

\newcommand{\Kepler}{\textit{Kepler}\xspace}

\newcommand{\emcee}{\texttt{emcee}\xspace}

\newcommand{\msini}{\ensuremath{M\sin{i}}\xspace}

\newcommand{\argperi}{\ensuremath{\omega}\xspace}
\newcommand{\ecosw}{\ensuremath{e \cos \argperi}\xspace}
\newcommand{\esinw}{\ensuremath{e \sin \argperi}\xspace}
\newcommand{\sqrtecosw}{\ensuremath{\sqrt{e} \cos \argperi}\xspace}
\newcommand{\sqrtesinw}{\ensuremath{\sqrt{e} \sin \argperi}\xspace}

\newcommand{\sigjit}[1]{
        \ensuremath{
                \ifthenelse{\equal{#1}{}}{\sigma_\mathrm{jit}}{\sigma_\mathrm{jit,#1}}}
        \xspace
}
\newcommand{\tp}{\ensuremath{T_{p}\xspace}}
\newcommand{\tc}{\ensuremath{T_{c}\xspace}}

\newcommand{\zfree}[1]{\ensuremath{\ifthenelse{\isempty{#1}}{z_{\mathrm{free}}^{*}}{z_{\mathrm{free,#1}}}}\xspace}

\newcommand{\radvel}{\texttt{RadVel}\xspace}

\newcommand{\radvelParameter}{\texttt{radvel.Parameter}\xspace}
\newcommand{\radvelParameters}{\texttt{radvel.Parameters}\xspace}
\newcommand{\radvelRVModel}{\texttt{radvel.RVModel}\xspace}
\newcommand{\radvelLikelihood}{\texttt{radvel.Likelihood}\xspace}
\newcommand{\radvelRVLikelihood}{\texttt{radvel.RVLikelihood}\xspace}
\newcommand{\radvelCompositeLikelihood}{\texttt{radvel.CompositeLikelihood}\xspace}
\newcommand{\radvelPosterior}{\texttt{radvel.Posterior}\xspace}

\newcommand{\radvelReport}{\texttt{radvel.RadvelReport}\xspace}
\newcommand{\radvelTable}{\texttt{radvel.TexTable}\xspace}

\newcommand{\xhat}{\ensuremath{\hat{\rm x}}\xspace}
\newcommand{\yhat}{\ensuremath{\hat{\rm y}}\xspace}
\newcommand{\zhat}{\ensuremath{\hat{\rm z}}\xspace}

\begin{document}

\title{RadVel: The Radial Velocity Modeling Toolkit}
\author{
Benjamin J.\ Fulton\altaffilmark{1,2,3},
Erik A.\ Petigura\altaffilmark{1,4},
Sarah Blunt\altaffilmark{1},
Evan Sinukoff\altaffilmark{1,2,5}
}

\altaffiltext{1}{California Institute of Technology, Pasadena, California, U.S.A.}
\altaffiltext{2}{Institute for Astronomy, University of Hawai`i at Ma\={a}noa, Honolulu, HI 96822, USA} 
\altaffiltext{3}{Texaco Fellow}
\altaffiltext{4}{Hubble Fellow}
\altaffiltext{5}{Natural Sciences and Engineering Research Council of Canada Graduate Student Fellow}

\begin{abstract}
\radvel is an open source Python package for modeling Keplerian orbits in radial velocity (RV) time series. \radvel provides a convenient framework to fit RVs using maximum a posteriori optimization and to compute robust confidence intervals by sampling the posterior probability density via Markov Chain Monte Carlo (MCMC). \radvel allows users to float or fix parameters, impose priors, and perform Bayesian model comparison. We have implemented realtime MCMC convergence tests to ensure adequate sampling of the posterior. \radvel can output a number of publication-quality plots and tables. Users may interface with \radvel through a convenient command-line interface or directly from Python. The code is object-oriented and thus naturally extensible. We encourage contributions from the community. Documentation is available at http://radvel.readthedocs.io.
\end{abstract}

\section{Introduction}
\label{sec:intro}

The radial velocity (RV) technique was among the first techniques to permit the discovery and characterization of extrasolar planets \citep[e.g.,][]{Campbell88, Latham89, Mayor95, Marcy96}. Prior to NASA's \Kepler Space Telescope \citep{Borucki10}, the RV technique accounted for the vast majority of exoplanet detections \citep{Akeson13}.

In the post-\Kepler era, the RV field has shifted somewhat from discovery to follow up and characterization of planets discovered by transit surveys. In the case of planets discovered via either technique the need to model RV orbits to extract planet masses (or minimum masses, \msini) is a critical tool necessary to understand the compositions and typical masses of exoplanets.

Several software packages have been written to address the need of the exoplanet community to fit RV time series.\footnote{a non-exhaustive list includes \texttt{RVLIN} \citep{WrightHoward09}, \texttt{Systemic} \citep{Meschiari09, Meschiari10}, \texttt{EXOFAST} \citep{Eastman13}, \texttt{rvfit} \citep{Iglesias15}, and \texttt{ExoSOFT} \citep{Mede17}} In designing \radvel, we have emphasized ease of use, flexibility, and extensibility. \radvel\ can be installed and a simple planetary system can modeled from the command-line in seconds. \radvel\ also provides an extensive and well-documented application programming interface (API) to perform complex fitting tasks. We employ modern Markov Chain Monte Carlo (MCMC) sampling techniques and robust convergence criteria to ensure accurately estimated orbital parameters and their associated uncertainties.

The goal of this paper is to document the core features of \radvel version 1.0 \citep{radvelcite}. Due to the evolving nature of \radvel, the most up to date documentation can be found at http://radvel.readthedocs.io (RTD page hereafter). This paper complements that documentation and is structured as follows: we describe the parameters involved to describe an RV orbit in Section \ref{sec:rv}, in Section \ref{sec:bayes} we discuss Bayesian inference as implemented in \radvel. The design of the code is described in Section \ref{sec:design} and the model fitting procedure is described in Section \ref{sec:fitting}. We explain how to install \radvel and walk through two example fits to demonstrate how the code is run in Section \ref{sec:examples}. We describe the mechanism for support and contributions in Section \ref{sec:community} and close with some concluding remarks in Section \ref{sec:conclusion}.

\section{The Radial Velocity Orbit}
\label{sec:rv}

\begin{figure*}[ht]
    \centering
    \includegraphics[scale=0.3]{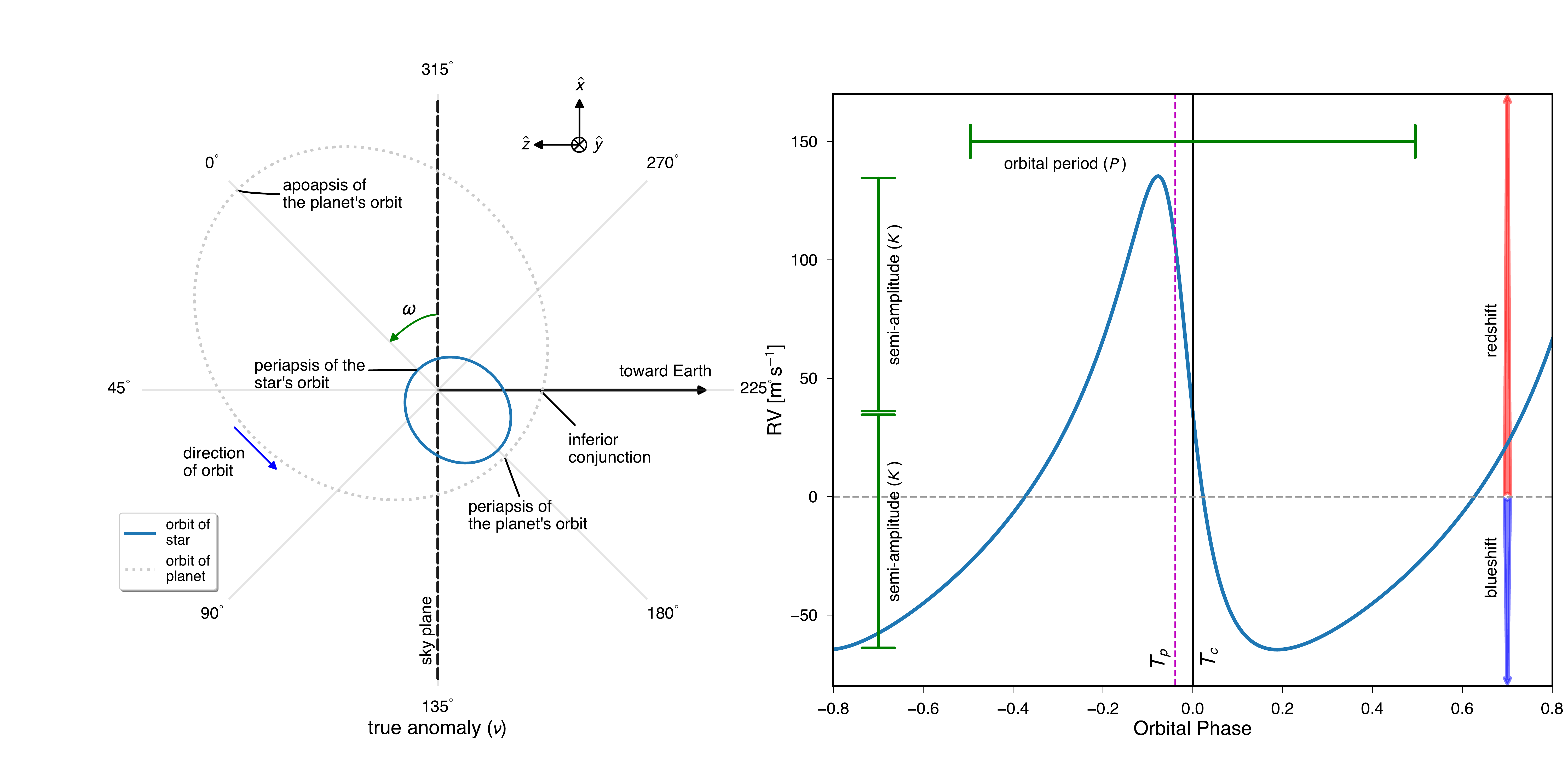}
    \caption{Diagram of a Keplerian orbit. \emph{Left:} Top-down view of an eccentric Keplerian orbit with relevant parameters labeled. The planet's orbit is plotted as a light grey dotted line and the star's orbit is plotted in blue (scale exaggerated for clarity). \emph{Right:} Model radial velocity curve for the same orbit with the relevant parameters labeled. We define a cartesian coordinate system described by the \xhat, \yhat, and \zhat unit vectors such that \zhat points away from the observer and \yhat is normal to the plane of the planet's orbit. In this coordinate system the x-y plane defines the sky plane. The plane of the orbit lies in the plane of the page.}
    \label{fig:orbit}
\end{figure*}

RV orbits are fundamentally described with five orbital elements: orbital period ($P$), a parameter that describes the orbital phase at a given time (we use the time of inferior conjunction, $T_c$), orbital eccentricity ($e$), the argument of periastron of the star's orbit ($\omega$), and the velocity semi-amplitude ($K$). We also include terms for the mean center-of-mass velocity ($\gamma$), plus linear ($\dot{\gamma}$) and quadratic ($\ddot{\gamma}$) acceleration terms in the RV model. Since RV measurement uncertainties generally do not take into account contributions from astrophysical and instrumental sources of noise we also fit for a ``jitter'' term ($\sigma$), which is added in quadrature with the measurement uncertainties. These parameters are listed and described in Table \ref{tab:orbit} for reference. Figure \ref{fig:orbit} depicts an example eccentric Keplerian orbit with several of these parameters annotated. We define a cartesian coordinate system described by the $\hat{\rm x}$, $\hat{\rm y}$, and $\hat{\rm z}$ unit vectors such that $\hat{\rm z}$ points away from the observer and $\hat{\rm y}$ is normal to the plane of the planet's orbit. In this coordinate system, the x-y plane defines the sky plane.

\begin{deluxetable}{lC}
\tablecaption{Keplerian Orbital Elements}
\tablehead{\colhead{Parameter Description} & \colhead{Symbol} }
\startdata
\sidehead{{\bf Keplerian Orbital Parameters}}
orbital period  &  $P$ \\
time of inferior conjunction (or transit)\tablenotemark{1}  &  $T_c$ \\
time of periastron\tablenotemark{1,2} & $T_p$ \\
eccentricity & $e$ \\
argument of periapsis of the star's orbit\tablenotemark{2} &  \omega  \\
velocity semi-amplitude  &  K \\
\hline
\sidehead{{\bf Mean and Acceleration Terms}}
mean center-of-mass velocity\tablenotemark{3}  &  \gamma_i  \\
linear acceleration term  &  \dot{\gamma} \\
second order acceleration term  &  \ddot{\gamma} \\
\hline
\sidehead{{\bf Noise Parameters}}
radial velocity ``jitter'' (white noise)\tablenotemark{3}  &  \sigma_i 
\enddata
\label{tab:orbit}
\tablenotetext{1}{Either $T_c$ or $T_p$ can be used to describe the phase of the orbit. Both are not needed simultaneously.}
\tablenotetext{2}{undefined for circular orbits}
\tablenotetext{3}{usually specific to each instrument ($i$)}
\end{deluxetable}

\subsection{Keplerian Solver}
\label{sec:kepler}
Synthesizing radial velocities involves solving the following system of equations,
\begin{eqnarray}
M & = & E - e \sin E \\
\nu & = & 2 \tan^{-1} \left( \sqrt{\frac{1 + e}{1 - e}} \tan\frac{E}{2}\right) \\
\dot{\rm z} & = & v_r = K [ \cos(\nu + \omega) + e \cos(\omega) ]
\end{eqnarray}
where $M$ is the mean anomaly, $E$ is the eccentric anomaly, and $e$ is the orbital eccentricity, $\nu$ is commonly referred to as the true anomaly, $K$ is the velocity semi-amplitude, and $v_r$ is the star's reflex RV caused by a single orbiting planet. Equation 1 is known as Kepler's equation and we solve it using the iterative method proposed by \cite{Danby88}, and reproduced in \cite{Murray99}.

Note that we equate $\dot{\rm z}$ to $v_r$ in our description of the RV orbit. In our coordinate system, the $\hat{\rm z}$ unit vector points \emph{away} from the observer. Historically, this has been the observational convention so that positive RV can be interpreted as a redshift of the star. However, some derivations of this equation in the literature \citep[e.g.,][]{Seager10, Deck14} define their coordinate system such that $\hat{\rm z}$ points \emph{toward} the observer and derive an equation for RV which is very similar to our Equation 3. The key difference when defining a coordinate system with $\hat{\rm z}$ pointing toward the observer is that the sign in Equation 3 must be flipped ($v_r = -\dot{\rm z}$) or, equivalently, $\omega$ must refer to the argument of periapsis of the \emph{planet's} orbit which is shifted by $\pi$ relative to the argument of periapsis of the star's orbit.

In the case of multiple planets the Keplerian orbits are summed together. We also add acceleration terms to account for additional perturbers in the system with orbital periods much longer than the observational baseline. The total RV motion of the star due to all companions ($\mathcal{V}_r$) is, 
\begin{equation}
\mathcal{V}_r = \sum_k^{N_{\rm pl}} v_{r,k} + \gamma + \dot{\gamma}(t - t_0) + \ddot{\gamma}(t - t_0)^2,
\label{eqn:vr}
\end{equation}
where $N_{\rm pl}$ is the total number of planets in the system and $t_0$ is an arbitrary abscissa epoch defined by the user.

\subsection{Parameterization}
In order to speed fitting convergence and avoid biasing parameters that must physically be finite and positive \citep[e.g.][]{Lucy71} analytical transformations of the orbital elements are often used to describe the orbit. In \radvel we have implemented several of these transformations into six different ``basis'' sets. One of the highlight features of \radvel is the ability to easily switch between these different bases. This allows the user to explore any biases that might arise based on the choice of parameterization, and/or impose priors on parameters or combinations of parameters that are not typically used to describe an RV orbit (e.g. a prior on \ecosw\ from a secondary eclipse detection). 

We have implemented the basis sets listed in Table \ref{tab:basis}.
Users can easily add additional basis sets by modifying the \texttt{radvel.basis.Basis} object. A string representation of the new basis should be added to the \texttt{radvel.basis.Basis.BASIS\_NAMES} attribute. The \texttt{radvel.basis.Basis.to\_synth} and \texttt{radvel.basis.Basis.from\_synth} methods should also be updated to properly transform the new basis to and from the existing ``synth'' basis.

\begin{deluxetable*}{ll}
\tablecaption{Parameterizations of the Orbit}
\tablehead{\colhead{Parameters describing orbit} & Notes }
\startdata
$P$, $\tp$, $e$, $\argperi$, $K$ & ``synth'' basis used to synthesize RVs \\
$P$, $\tc$, $\sqrtecosw$, $\sqrtesinw$, $K$ & standard basis for fitting and posterior sampling \\
$P$, $\tc$, $\sqrtecosw$, $\sqrtesinw$, $\ln{K}$ & forces $K>0$ \\
$P$, $\tc$, $\ecosw$, $\esinw$, $K$ & imposes a linear prior on $e$ \\
$P$, $\tc$, $e$, $\argperi$, $K$ & slower MCMC convergence \\
$\ln{P}$, $\tc$, $e$, $\argperi$, $\ln{K}$ & useful when $P$ is long compared to the observational baseline
\enddata
\label{tab:basis}
\end{deluxetable*}

Since priors are assumed to be uniform in the \emph{fitting} basis this imposes implicit priors on the Keplerian orbital elements. For example, choosing the ``$\ln{P}$, $\tc$, $e$, $\argperi$, $\ln{K}$'' basis would impose a prior that favors small $P$ and $K$ values since there is much more phase space for the MCMC chains to explore near $P = K = 0$. See \citet{Eastman13} for a detailed description of the implicit priors imposed on $e$ and $\omega$ based on the choice of fitting in \ecosw and \esinw, \sqrtecosw and \sqrtesinw, or $e$ and $\omega$ directly. In the case that the data does not constrain or weakly constrains some or all orbital parameters, the choice of the fitting basis becomes important due to these implicit priors. We usually prefer to perform fitting and posterior sampling using the $P$, $\tc$, $\sqrtecosw$, $\sqrtesinw$, $K$ basis since this imposes flat priors on all of the orbital elements, avoids biasing $K>0$, and helps to speed MCMC convergence. However, the $P$, $\tc$, $\sqrtecosw$, $\sqrtesinw$, $\ln{K}$, and $\ln{P}$, $\tc$, $e$, $\argperi$, $\ln{K}$ bases can be useful in some scenarios, especially when $K$ is large and $P$ is long compared to the observational baseline. There is no default basis. The choice of fitting basis must be made explicitly by the user. It is generally good practice to perform the fit in several different basis sets to check for consistency.

Inspection of Equation 3 shows that $v_r$ for $K=+q$ and $\omega=\lambda$ is identical to $v_r$ when $K=-q$ and $\omega=\lambda+\pi$. For low signal-to-noise detections where the MCMC walkers (see Section \ref{sec:mcmc}) can jump between the $K=+q$ and $K=-q$ solutions this degeneracy can lead to bimodal posterior distributions for $K$ reflected about $K=0$. The posterior distributions for $T_c$ and/or $\omega$ will also be bimodal. In these cases, we advise the user to proceed with caution when interpreting the posterior distributions and to explore a variety of basis sets and priors (see Section \ref{sec:priors}) to determine their impact on the resulting posteriors.


\section{Bayesian Inferrence}
\label{sec:bayes}

We model RV data following the standard practices of Bayesian inference. The goal is to infer the posterior probability density $p$ given a dataset ($\mathcal{D}$) and priors as in Bayes' Theorem:
\begin{equation}
p(\theta | \mathcal{D}) \propto \mathcal{L}(\theta | \mathcal{D})p(\theta).
\end{equation}
The Keplerian orbital parameters are contained in the $\theta$ vector. 
$\mathcal{L}(\theta | \mathcal{D})$ is the likelihood that the data is drawn from the model described by the parameter set $\theta$. Assuming Gaussian distributed noise, the likelihood is
\begin{equation}
\begin{split}
\ln{\mathcal{L}_i(\theta | \mathcal{D}_i)} = - \frac{1}{2} \sum_j\frac{(\mathcal{V}_{r,j}(t, \theta) - d_j )^2}{e_j^2 + \sigma_i^2} \\
- \ln{\sqrt{2\pi(e_j^2 + \sigma_i^2)}},
\end{split}
\end{equation}
where $\mathcal{V}_{r,j}$ is the Keplerian model (Equation \ref{eqn:vr}) predicted at the time ($t$) of each RV measurement ($d_j$), $e_j$ is the measurement uncertainty associated with each $d_j$, and $\sigma_i$ is a Gaussian noise term to account for any astrophysical or instrumental noise not included in the measurement uncertainties. The $\sigma_i$ terms are unique to each instrument ($i$). For datasets containing velocities from multiple instruments the total likelihood is the sum of the natural log of the likelihoods for each instrument.
\begin{equation}
\ln{\mathcal{L}(\theta | \mathcal{D})} = \sum_i{\ln{\mathcal{L}_i(\theta | \mathcal{D}_i)}}
\end{equation}
We sample the posterior probability density surface using MCMC (see Section \ref{sec:mcmc}). The natural log of the priors are applied as additional additive terms such that 
\begin{equation}
\ln{\mathcal{L}(\theta | \mathcal{D})}p(\theta) = \ln{\mathcal{L}(\theta | \mathcal{D})} + \sum_k \ln{\mathcal{P}_k(\theta)},
\end{equation}
for each prior ($\mathcal{P}_k$). If no priors are explicitly defined by the user then all priors are assumed to be uniform and $\sum_k \ln{\mathcal{P}_k(\theta)} = 0$.

\section{Code Design}
\label{sec:design}

\begin{figure*}
    \centering
    \includegraphics[scale=0.6]{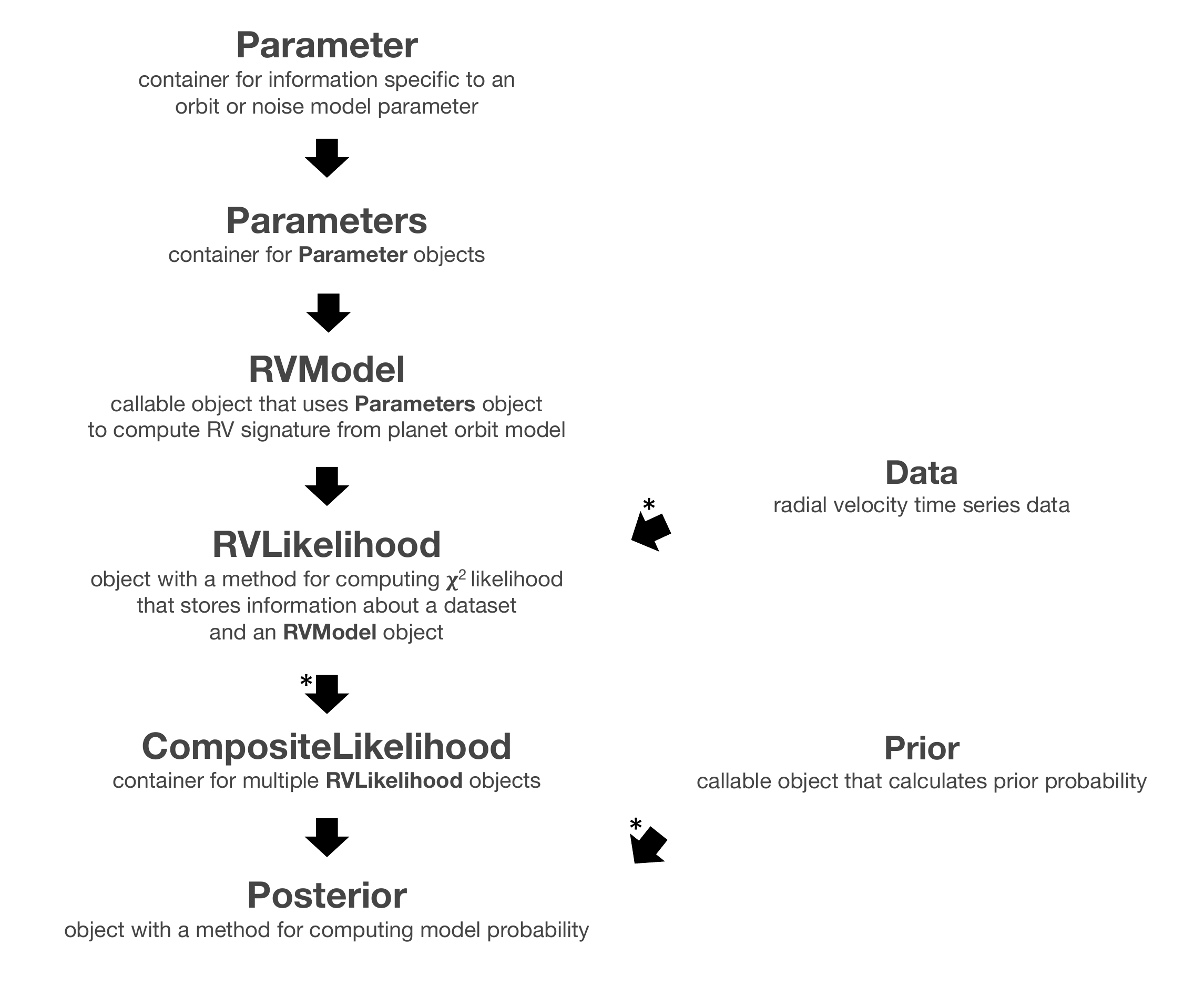}
    \caption{Class diagram for the \radvel package showing the relationships between the various objects contained within the \radvel package. Arrows point from attributes to their container objects (i.e. a \radvelParameter object is an attribute of a \radvelParameters object). Pertinent characteristics are summarized beneath each object. An asterisk next to an arrow indicates that the container object typically contains multiple attribute objects of the indicated type.}
    \label{fig:layout}
\end{figure*}
The fundamental quantity for model fitting and parameter estimation is the posterior probability density $p$, which is represented in \radvel as a Python object. Users create $p$ by specifying a likelihood $\mathcal{L}$, priors $\mathcal{P}$, model $\mathcal{V}_r$, and data $\mathcal{D}$, which are also implemented as objects. Here, we describe each of these objects, which are the building blocks of the \radvel API. Figure \ref{fig:layout} summarizes these objects and their hierarchy.

\subsection{Parameters}

The posterior probability density $p$ is a surface in $\mathcal{R}^N$, where $N$ is number of free parameters. We specify coordinates in this parameter space using a \radvelParameters object. \radvelParameters is a container object that inherits from Python's ordered dictionary object, \texttt{collections.OrderedDict}\xspace.
We modeled this dictionary representation after the \texttt{lmfit}\footnote{https://lmfit.github.io/lmfit-py/} \citep{Newville14} API which allows users to conveniently interface with variables via string keys (as opposed to integer indexes).
Auxiliary attributes, such as the number of planets and the fitting basis, are also stored in the \radvelParameters object outside of the dictionary representation.

Each element of the \radvelParameters dictionary is represented as a \radvelParameter object which contains the parameter value and a boolean attribute which specifies if the parameter is fixed or allowed to float.\footnote{Note that this representation is different from that used in \radvel versions $<1.0$.}

\subsection{Model Object}
The \radvelRVModel class is a callable object that computes the radial velocity curve that corresponds to the parameters stored in the \radvelParameters object. Calculating the model RV curve requires solving Kepler's equation (see Section \ref{sec:kepler}) and is computationally intensive.
This solver is implemented in C in order to maximize performance. We also provide a Python implementation so that users may run \radvel without compiling C code (albeit at slower speeds).

\subsection{Likelihood Object}
\label{sec:like}
The primary function of the \radvelLikelihood object is to establish the relationship between a model and the data. It is a generic class which is meant to be inherited by objects designed for specific applications such as the \radvelRVLikelihood object. Most fitting packages \citep[e.g. \emcee,][]{DFM13} require functions which take vectors of floating-point values as inputs and output a single goodness-of-fit metric. These conversions between the string-indexed \radvelParameters object and ordered arrays of floats containing only the parameters that are allowed to vary are handled within the \radvelLikelihood object.

The \radvelRVLikelihood object is a container for a single radial velocity dataset (usually from a single instrument) and a \radvelRVModel object. The \texttt{radvel.Likelihood.logprob} method returns the natural log of the likelihood of the model evaluated at the parameter values contained in the \texttt{params} attribute given the data contained in the \texttt{x}, \texttt{y}, and \texttt{yerr} attributes. The \texttt{extra\_params} attribute contains additional parameters that are not needed to calculate the Keplerian model, but are needed in the calculation of the likelihood (e.g. jitter).

The \radvelCompositeLikelihood object is simply a container for multiple \radvelRVLikelihood objects which is constructed in the case of multi-instrument datasets. The \texttt{logprob} method of the \radvelCompositeLikelihood adds the results of the \texttt{logprob} methods for all of the \radvelRVLikelihood objects contained within the \radvelCompositeLikelihood.

\newpage
\subsection{Posterior Object}
The \radvelPosterior object is very similar to the \radvelLikelihood object but it also contains any user-defined priors.
The \texttt{logprob} method of the \radvelPosterior object is then the natural log of the likelihood of the data given the model and priors.

\subsection{Priors}
\label{sec:priors}
Priors are defined in the \texttt{radvel.prior} module and should be callable objects which return a single value to be multiplied by the likelihood. Several useful example priors are already implemented. 
\begin{itemize}
\item \texttt{EccentricityPrior} can be used to set upper limits on the planet eccentricities.
\item \texttt{GaussianPrior} can be used to assign a prior to a parameter value with a given center ($\mu$) and width ($\sigma$). 
\item \texttt{PositiveKPrior} can be used to force planet semi-amplitudes to be positive.\footnote{This should be used with extreme caution to avoid biasing results toward non-zero planet masses \citep{Lucy71}.} 
\item \texttt{HardBounds} prior is used to impose strict limits on parameter values.
\end{itemize}
Other priors are continuously being implemented and we encourage users to frequently check the API documentation on the RTD page for new priors.

\section{Model-fitting}
\label{sec:fitting}

\subsection{Maximum a Posteriori Fitting}
The set of orbital parameters which maximizes the posterior probability (maximum a posteriori optimization, MAP) are found using Powell's method \citep{Powell64} as implemented in \texttt{scipy.optimize.minimize}.\footnote{Any of the optimization methods implemented in \texttt{scipy.optimize.minimize} which do not require pre-calculation of derivatives (e.g. Nelder-Mead) can be swapped in with a simple modification to the code in the \texttt{radvel.fitting} module.} The code that performs the minimization can be found in the \texttt{radvel.fitting} submodule.

\subsection{Uncertainty Estimation}
\label{sec:mcmc}

The \texttt{radvel.mcmc} module handles the MCMC exploration of the posterior probability surface ($p$) in order to estimate parameter uncertainties. We use the MCMC package \emcee \citep{DFM13}, which employs an Affine Invariant sampler \citep{Goodman10}.
The MCMC sampling generally explores a fairly wide range of parameter values but \radvel should not be treated as a planet discovery tool. The \radvel MCMC functionality is simply meant to determine the size and shape of the posterior probability density.

It is important to check that all of the independent walkers in the MCMC chains have adequately explored the same maximum on the posterior probability density surface and are not stuck in isolated local maxima. The Gelman-Rubin statistic \citep[G-R,][]{Gelman03} is one metric to check for ``convergence'' by comparing the intra-chain variances to the inter-chain variances. G-R values close to unity indicate that the chains are converged.

We initially run a set of MCMC chains until the G-R statistic is less than 1.03 for all free parameters as a burn-in phase. These initial steps are the discarded and new chains are launched from the last positions. We note that this is a perticularily conservative approach to burn-in that is enabled thanks to the very fast Keplerian model calculation and convergence of \radvel. Users may relax the G-R$<1.03$ burn-in requirement via command line flags or in the arguments of the \texttt{radvel.mcmc.mcmc} function. After the burn-in phase we follow the prescription of \citet{Eastman13} to check the MCMC chains for convergence after every 50 steps. The chains are deemed well-mixed and the MCMC run is halted when the G-R statistic is less than 1.01 and the number of independent samples \citep[$T_{z}$ statistic,][]{Ford06} is greater than 1000 for all free parameters for at least 5 consecutive checks. We note that these statistics can not be calculated between walkers within an ensemble so we calculate G-R and $T_{z}$ across completely independent ensembles of samplers. By default we run 8 independent ensembles in parallel with 50 walkers per ensemble for up to a maximum of 10000 steps per walker or until convergence is reached. 

These defaults can be customized on the command line or in the arguments of the \texttt{radvel.mcmc.mcmc} function. Each of the independent ensembles are run on separate CPUs so the number of ensembles should not exceed the number of CPUs available or significant slowdown will occur. Default initial step sizes for all free parameters are set to 10\% of their value except period which is set to 0.001\% of the period. These initial step sizes can be customized by setting the \texttt{mcmcscale} attributes of the \radvelParameter objects.

\section{Examples}
\label{sec:examples}
Users interact with \radvel either through a command-line interface (CLI) or through the Python API. Below, we run through an example RV analysis of HD 164922 from \citep{Fulton16} using the CLI. The Python API exposes additional functionality that may be used for special case fitting and is described in the advanced usage page on the RTD website.

\subsection{Installation}
\label{sec:install}

To install \radvel, users must have working version of Python 2 or 3.\footnote{As the community transitions from Python 2 to 3, we will likely drop support for Python 2.} We recommend the Anaconda Python distribution.\footnote{https://www.anaconda.com}

Users may then install \radvel from the Python Package Index (pip).\footnote{https://pypi.python.org/pypi}$^{,}$\footnote{Early \radvel adopters may see installation conflicts with
early versions of \radvel. We recommend manually removing old versions and performing a fresh install with \texttt{pip}.}
{\scriptsize 
\begin{lstlisting}[language=bash]
  $ pip install radvel
\end{lstlisting}}

\subsection{Command-Line Interface}
\subsubsection{Setup Files}
\label{sec:setup}

The setup file is the central component in the command-line interface execution of \radvel. This file is a Python script which defines the number of planets in the system, initial parameter guesses, stellar parameters if present, priors, reads and defines datasets, initializes a model object, and defines metadata associated with the run. One setup file should be produced for each planetary system that the user attempts to model using the command-line interface.
Two example setup files can be downloaded from the GitHub repo. A complete example is provided below with inline comments to describe the various components.

\begin{lstlisting}[language=Python]
# Required packages for setup
import os
import pandas as pd
import numpy as np
import radvel as rv

# name of the star used for plots and tables
starname = 'HD164922'

# number of planets in the system
nplanets = 2 

# list of instrument names. Can be whatever 
# you like but should match 'tel' column in the 
# input data file.
instnames = ['k', 'j', 'a']

# number of instruments with unique
# velocity zero-points
ntels = len(instnames)

# Fitting basis, see rv.basis.BASIS_NAMES
# for available basis names
fitting_basis = 'per tc secosw sesinw k'

# reference epoch for plotting purposes
bjd0 = 2450000.

# (optional) customize letters 
# corresponding to planet indicies
planet_letters = {1: 'b', 2:'c'}

# Define prior centers (initial guesses) in a basis
# of your choice (need not be in the fitting basis)
# initialize Parameters object
params = rv.Parameters(nplanets, 
                    basis='per tc e w k')

# period of 1st planet
params['per1'] = rv.Parameter(value=1206.3)    
# time of inferior conjunction of 1st planet
params['tc1'] = rv.Parameter(value=2456779.)  
# eccentricity
params['e1'] = rv.Parameter(value=0.01)   
# argument of periastron of the star's orbit 
params['w1'] = rv.Parameter(value=np.pi/2)      
# velocity semi-amplitude
params['k1'] = rv.Parameter(value=10.0)
 
# same parameters for 2nd planet ...
params['per2'] = rv.Parameter(value=75.771)    
params['tc2'] = rv.Parameter(value=2456277.6)
params['e2'] = rv.Parameter(value=0.01)
params['w2'] = rv.Parameter(value=np.pi/2)
params['k2'] = rv.Parameter(value=1.0)

# slope and curvature
params['dvdt'] = rv.Parameter(value=0.0)        
params['curv'] = rv.Parameter(value=0.0)

# zero-points and jitter terms for each instrument
params['gamma_j'] = rv.Parameter(1.0)
params['jit_j'] = rv.Parameter(value=2.6)


# Convert input orbital parameters into the
# fitting basis
params = params.basis.to_any_basis(
                    params,fitting_basis)

# Set the 'vary' attributes of each of the parameters
# in the fitting basis. A parameter's 'vary' attribute
# should be set to False if you wish to hold it fixed
# during the fitting process. By default, all 'vary'
# parameters are set to True.
params['dvdt'].vary = False
params['curv'].vary = False

# Load radial velocity data, in this example the
# data are contained in a csv file, the resulting
# dataframe or must have 'time', 'mnvel', 'errvel',
# and 'tel' keys the velocities are expected
# to be in m/s
path = os.path.join(rv.DATADIR, '164922_fixed.txt')
data = pd.read_csv(path, sep=' ')

# Define prior shapes and widths here.
priors = [
    # Keeps eccentricity < 1 for all planets
    rv.prior.EccentricityPrior( nplanets ),  
    # Keeps K > 0 for all planets         
    rv.prior.PositiveKPrior( nplanets ),   
    # Hard limits on jitter parameters
    rv.prior.HardBounds('jit_j', 0.0, 10.0),
    rv.prior.HardBounds('jit_k', 0.0, 10.0),
    rv.prior.HardBounds('jit_a', 0.0, 10.0)
]

# abscissa for slope and curvature terms
# (should be near mid-point of time baseline)
time_base = 0.5*(data.time.min() + data.time.max())  

# optional argument that can  contain stellar mass
# in solar units (mstar) and uncertainty (mstar_err).
# If not set, mstar will be set to nan.
stellar = dict(mstar=0.874, mstar_err=0.012)
\end{lstlisting}

\subsubsection{Workflow}
\label{sec:cli}

The \radvel\ CLI is provided for the convenience of the user and is the standard operating mode of \radvel. It acts as a wrapper for much of the underlying API. Most users will likely find this to be the easiest way to run \radvel\ fits and produce the standard outputs.

Here we provide a walkthrough for a basic \radvel\ fit for a multi-planet system with RV data collected using three different instruments. The data for this example is taken from \citet{Fulton16}. The first step is to create a setup file for the fit. In this case, we have provided a setup file in the GitHub repo (\texttt{example\_planets/HD164922.py}). Once we have a setup file we always need to run a MAP fit first. This is done using the \texttt{radvel fit} command.
{\scriptsize 
\begin{lstlisting}[language=bash]
  $ radvel fit -s /path/to/HD164922.py
\end{lstlisting}}
This command will produce some text output summarizing the result which shows the final parameter values after the MAP fit.
A new directory will be created in the current working directory with the same name of the setup file. In this case, a directory named \texttt{HD164922} was created and it contains a \texttt{HD164922\_post\_obj.pkl} file and a \texttt{HD164922\_radvel.stat} file. Both of these files are used internally by \radvel to keep track of the components of the fit that have already been run. All of the output associated with this \radvel\ fit will be put in this directory.

It is useful to inspect the MAP fit using the \texttt{radvel plot} command.
{\scriptsize 
\begin{lstlisting}[language=bash]
  $ radvel plot -t rv -s /path/to/HD164922.py
\end{lstlisting} 
}
\begin{figure*}
    \centering
    \includegraphics[scale=0.7]{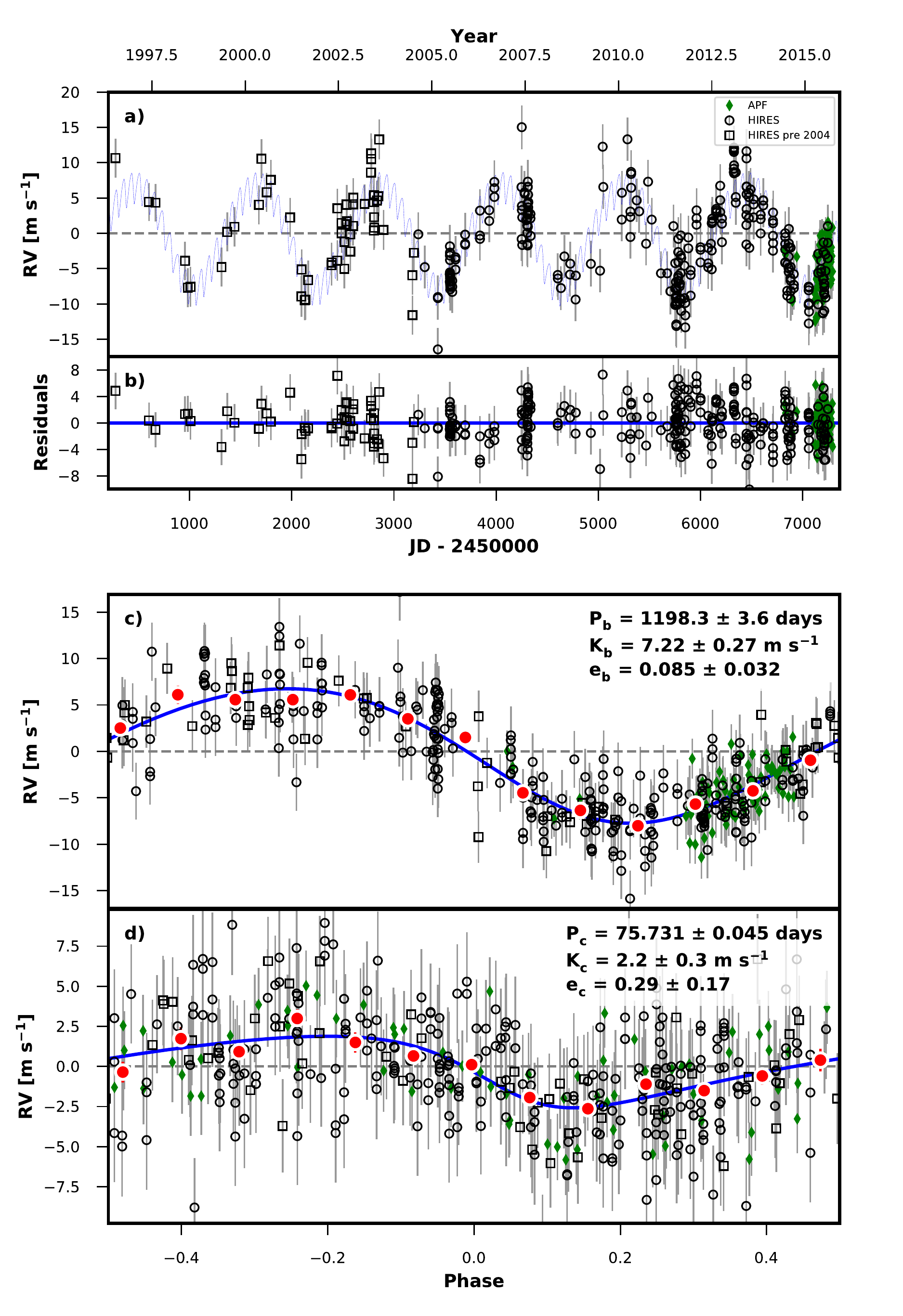}
    \caption{Example RV time series plot produced by \radvel for the HD 164922 fit with two planets and three spectrometers (see Section \ref{sec:cli}).
\noindent This shows the MAP 2-planet Keplerian orbital model.
The blue line is the 2-planet model. The RV jitter terms are included in the plotted measurement uncertainties.
{\bf b)} Residuals to the best fit 2-planet model.
{\bf c)} RVs phase-folded to the ephemeris of planet b. The Keplerian orbital models for all other planets have been subtracted.
The small point colors and symbols are the same as in panel {\bf a}.
Red circles are the same velocities binned in 0.08 units of orbital phase.
The phase-folded model for planet b is shown as the blue line.
Panel {\bf d)} is the same as panel {\bf c)} but for planet HD 164922 c.}
    \label{fig:rv}
\end{figure*}
\noindent This will produce a new PDF file in the output directory called \texttt{HD164922\_rv\_multipanel.pdf} that looks very similar to Figure \ref{fig:rv}. The annotated parameter uncertainties will only be printed if MCMC has already been run. The \texttt{-t rv} flag tells \radvel\ to produce the standard RV time series plot. Several other types of plots can also be produced using the \texttt{radvel plot} command.

If the fit looks good then the next step is to run an MCMC exploration to estimate parameter uncertainties.
{\scriptsize 
\begin{lstlisting}[language=bash]
  $ radvel mcmc -s /path/to/HD164922.py
\end{lstlisting} 
}
\noindent Once the MCMC chains have converged or the maximum step limit is reached (see Section \ref{sec:mcmc}) two additional output files will be produced: \texttt{HD164922\_chains.csv.tar.bz2} and \texttt{HD164922\_post\_summary.csv}. The \texttt{chains.csv.tar.bz2} file contains each step in the MCMC chains. All of the independent ensembles and walkers have been combined together such that there is one column per free parameter. 
The \texttt{post\_summary.csv} file contains the median and 68th percentile credible intervals for all free parameters.

\begin{figure*}
    \centering
    \includegraphics[scale=0.21]{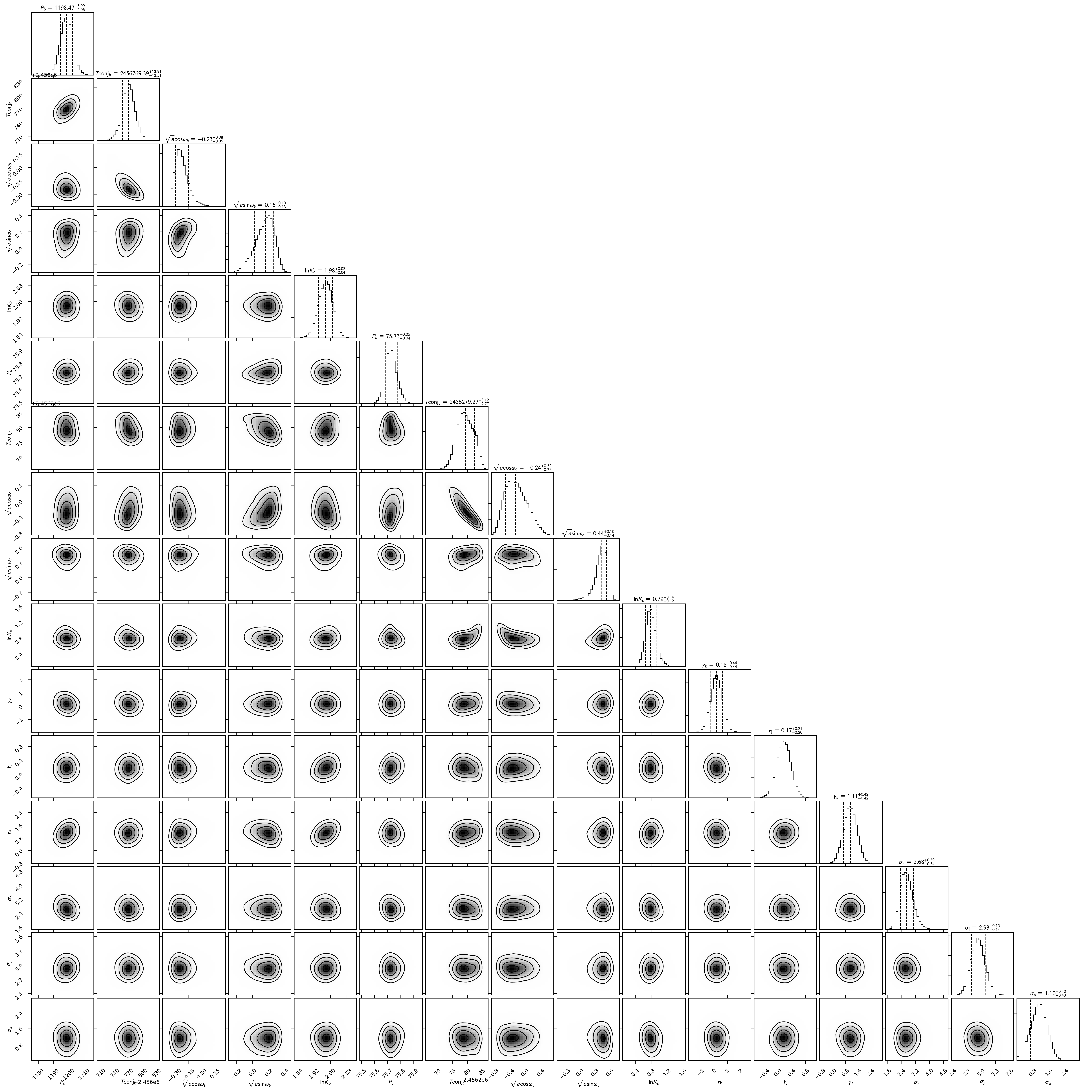}
    \caption{Corner plot showing all joint posterior distributions derived from the MCMC sampling. Histograms showing the marginalized posterior distributions for each parameter are also included. This plot is produced as part of the CLI example for HD 164922 as discussed in Section \ref{sec:cli}.}
    \label{fig:corner}
\end{figure*}

\begin{figure*}[ht]
    \centering
    \includegraphics[scale=0.6]{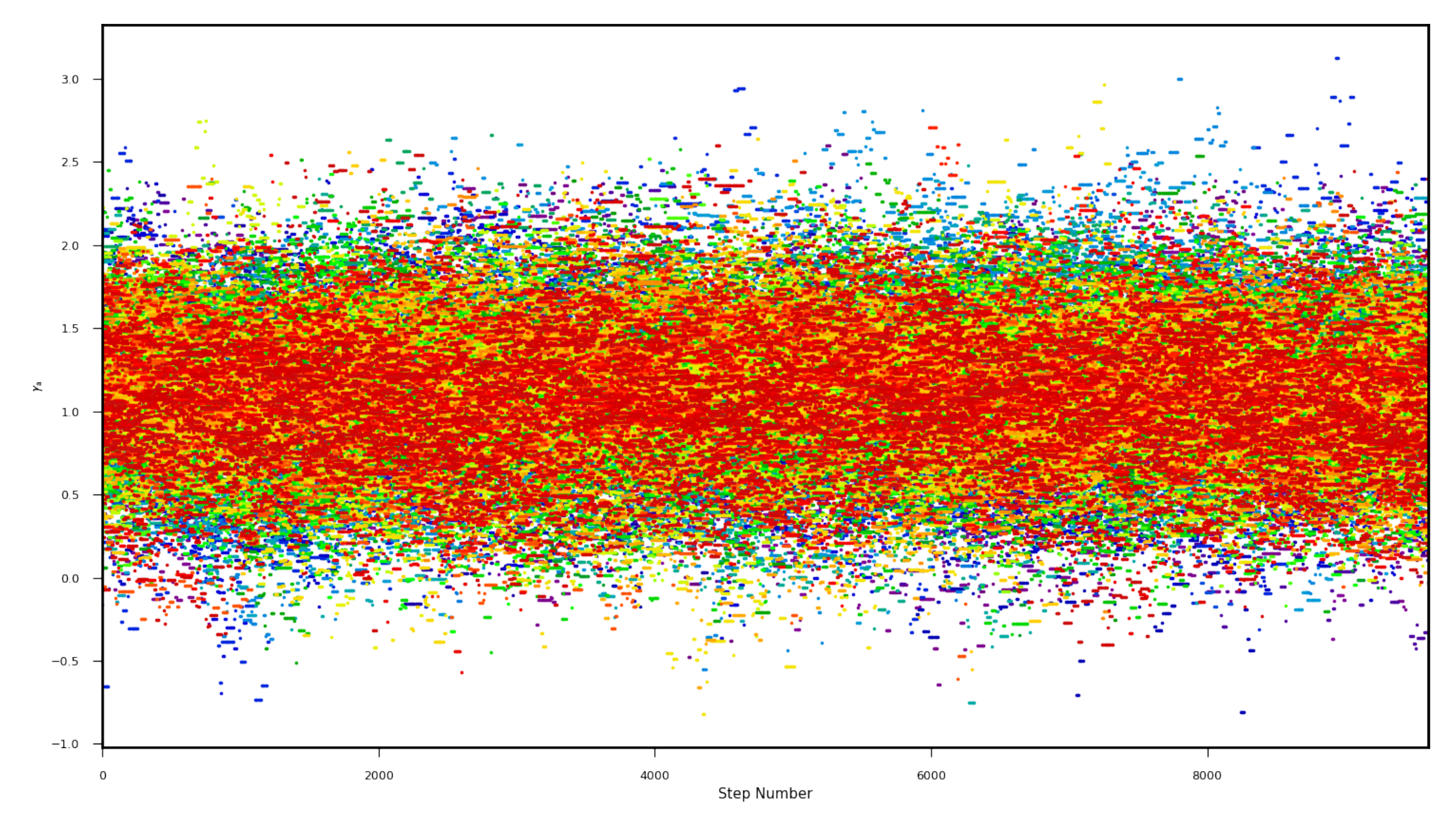}
    \caption{Trend plot produced by \radvel\ showing parameter evolution during the MCMC exploration. This plot is produced as part of the CLI example for HD 164922 as discussed in Section \ref{sec:cli}. As an example, we show only a single page of the PDF output. In practice, there will be one plot like this for each free parameter. Each independent ensemble and walker is plotted in different colors. If the MCMC runs have successfully converged these plots should look like white noise will all colors mixed together. Large-scale systematics and/or single chains (colors) isolated from the rest likely indicate problems with the initial parameter guesses and/or choices of priors.}
    \label{fig:trend}
\end{figure*}

Now that the MCMC has finished we can make some additional plots.
{\scriptsize 
\begin{lstlisting}[language=bash]
  $ radvel plot -t rv -s /path/to/HD164922.py
  $ radvel plot -t corner -s /path/to/HD164922.py
  $ radvel plot -t trend -s /path/to/HD164922.py
\end{lstlisting} 
}
\noindent The RV time series plot (Figure \ref{fig:rv}) now includes the parameter uncertainties in the annotations. In addition, a ``corner'' or triangle plot showing all parameter covariances produced by the \texttt{corner} Python package \citep[Figure \ref{fig:corner}, ][]{DFM16}, and a ``trend'' plot showing the evolution of the parameter values as they step through the MCMC chains will be produced (Figure \ref{fig:trend}).

In this case we have also defined the stellar mass (\texttt{mstar}) and uncertainty (\texttt{mstar\_err}) in the \texttt{stellar} dictionary in the setup file. This allows us to use the \texttt{radvel derive} command to convert the velocity semi-amplitude ($K$), orbital period ($P$), and eccentricity ($e$) into a minimum planet mass (\msini).
{\scriptsize 
\begin{lstlisting}[language=bash]
  $ radvel derive -s /path/to/HD164922.py
  $ radvel plot -t derived -s /path/to/HD164922.py
\end{lstlisting} 
}
This produces the file \texttt{HD164922\_derived.csv.tar.bz2} in the output directory which contains columns for each of the derived parameters. For the case of transiting planets planetary radii can also be specified (see the \texttt{epic203771098.py} example file) to allow the computation of planet densities. Synthetic Gaussian posterior distributions are created for these stellar parameters and these synthetic posteriors are multiplied by the real posterior distributions in order to properly account for the uncertainties in both the stellar and orbital parameters. A corner plot for the derived parameters is created using the \texttt{radvel plot -t derived} command.

An optional model comparison table (Table \ref{tab:comp}) can be created using the \texttt{radvel bic} command.
{\scriptsize 
\begin{lstlisting}[language=bash]
  $ radvel bic -t nplanets -s /path/to/HD164922.py
\end{lstlisting} 
}
\noindent This produces a table summarizing model comparisons with 0 to $N_{\rm pl}$ planets where $N_{\rm pl}$ is the number of planets in the system specified in the setup file. This allows the user to compare models with fewer planets to ensure that their adopted model is statistically favored. The comparisons are performed by fixing the jitter parameters to their MAP values from the full $N_{\rm pl}$ planet fit.

\radvel can also produce publication-quality plots, tables, and a summary report in \latex\ and PDF form. The functionality contained in the \radvelReport and \radvelTable objects depend on the existence of a setup file (see Section \ref{sec:setup}) which is usually only present when utilizing the CLI. \radvel reports contain \latex\ tables showing the MAP values and credible intervals for all orbital parameters, a summary of non-uniform priors, and a model comparison table.\footnote{only if the \texttt{radvel bic} command has been run} A RV time series plot and a corner plot are also included. If stellar and/or planetary parameters are specified in the setup file (see Section \ref{sec:setup}) then a second corner plot is included with the derived parameters including planet masses (\msini) and densities (if transiting and planet radii given).

The command, 
{\scriptsize 
\begin{lstlisting}[language=bash]
  $ radvel report -s /path/to/HD164922.py
\end{lstlisting} 
}
\noindent creates the \latex\ file \texttt{HD164922\_results.tex} in the output directory and compiles it into a PDF using the \texttt{pdflatex} command.\footnote{The \texttt{pdflatex} binary is assumed to be in the system's path by default, but the full path to the binary may be specified using the \texttt{-{}-latex-compiler} flag if it installed in a non-standard location.} 
The summary PDF is a great way to send \radvel\ results to colleagues and since the \latex\ code for the report is also saved the tables can be copied directly into a manuscript to avoid transcription errors.


\begin{deluxetable*}{lrrr}
\tablecaption{Model Comparison}
\tablehead{\colhead{Statistic} & \colhead{0 planets} & \colhead{1 planets} & \colhead{{\bf 2 planets (adopted)}}}
\startdata
$N_{\rm data}$ (number of measurements)  & 401 & 401 & 401\\
$N_{\rm free}$ (number of free parameters)  & 3 & 8 & 13\\
RMS (RMS of residuals in m s$^{-1}$)  & 4.97 & 3.01 & 2.91\\
$\chi^{2}$ (jitter fixed)  & 1125.56 & 431.7 & 397.29\\
$\chi^{2}_{\nu}$ (jitter fixed)  & 2.83 & 1.1 & 1.02\\
$\ln{\mathcal{L}}$ (natural log of the likelihood)  & -1356.54 & -1009.61 & -992.41\\
BIC (Bayesian information criterion)  & 2731.06 & 2067.17 & 2062.74
\enddata
\label{tab:comp}
\end{deluxetable*}

\section{RadVel and the Community}
\label{sec:community}
\subsection{Support}

Please report any bugs or problems to the issues tracker on the GitHub repo page.\footnote{https://github.com/California-Planet-Search/radvel/issues}
Bugfixes in response to issue reports will be included in the next version release and will be available by download from PyPI (via \texttt{pip install radvel -{}-upgrade}) at that time.

\subsection{Contributing Code}
The \radvel\ codebase is open source and we encourage contributions from the community. All development will take place on GitHub. Developers should fork the repo and submit pull requests into the \texttt{next-release} base branch. These pull requests will be reviewed by the maintainers and merged into the \texttt{master} branch to be included in the next tagged release. We ask that developers follow these simple guidelines when contributing code.
\begin{itemize}
\item Do not break any existing functionality. This will automatically be checked when a pull request is created via Travis Continuous Integration\footnote{https://travis-ci.org}.
\item Develop with support for Python 3 and backwards-compatibility with Python 2 where possible
\item Code according to PEP8 standards\footnote{https://www.python.org/dev/peps/pep-0008/}
\item Document your code using Napolean-compatible docstrings\footnote{https://sphinxcontrib-napoleon.readthedocs.io} 
following the Google Python Style Guide\footnote{http://google.github.io/styleguide/pyguide.html}
\item Include unit tests that touch any new code using the \texttt{nosetests} framework\footnote{http://pythontesting.net/framework/nose}. Example unit tests can be found in the \texttt{radvel/radvel/tests} subdirectory of the repo.
\end{itemize}

\subsection{Future Work}
\label{sec:future}

\radvel is currently under active development and will grow to incorporate the modeling needs of the exoplanet community. Specific areas for future include:
\begin{itemize}
\item {\em Gaussian Process noise modeling}. Implement likelihoods that incorporate Gaussian process descriptions of RV variability
\end{itemize}


The authors have several potential improvements to
\radvel in mind or currently under development. We encourage
the community to suggest other wishlist items
or contribute insight and/or code to ongoing development
efforts using the GitHub issue tracker. Our current
wishlist is as follows:
\begin{itemize}
\item Improve performance of the dictionary key-based string indexing
\item Include other types of model comparisons in the \texttt{radvel bic} command (e.g. eccentric vs. circular fits)
\item Add ability to simultaneously fit other datasets (e.g. transit photometry, transit timing variations, astrometry)
\end{itemize}

\section{Conclusion}
\label{sec:conclusion}
We have provided a flexible, open-source toolkit for modeling RV data written in object-oriented Python. The package is designed to model the RV orbits of systems with multiple planets and data collected from multiple instruments. It features a convenient command-line interface and a scriptable API. \radvel utilizes a fast Keplerian solver written in C and robust, real-time convergence checking of the MCMC chains. It supports multiple parameterizations of the RV orbit and contains convenience functions for converting between parameterizations. \radvel\ has already been used to model RV orbits in at least nine refereed publications.\footnote{\citep{Sinukoff17a, Sinukoff17b, Petigura17b, Crossfield17, Weiss17, Grunblatt17, Christiansen17, Teske17}.} In addition, \citet{Teske17} demonstrated that \radvel produces results consistent with the results of the \texttt{Systemic} RV fitting package \citep{Meschiari09, Meschiari10}. We encourage the community to continue using \radvel\ for their RV modeling needs and to contribute to it's future development.

\acknowledgments{
EAP acknowledges support from Hubble Fellowship grant HST-HF2-51365.001-A awarded by the Space Telescope Science Institute, which is operated by the Association of Universities for Research in Astronomy, Inc. for NASA under contract NAS 5-26555. 
ES is supported by a post-graduate scholarship from the Natural Sciences and Engineering Research Council of Canada.
}

\bibliographystyle{aasjournal}
\bibliography{references}

\begin{thebibliography}{}
\expandafter\ifx\csname natexlab\endcsname\relax\def\natexlab#1{#1}\fi
\providecommand{\url}[1]{\href{#1}{#1}}

\bibitem[{{Akeson} {et~al.}(2013){Akeson}, {Chen}, {Ciardi}, {Crane}, {Good},
  {Harbut}, {Jackson}, {Kane}, {Laity}, {Leifer}, {Lynn}, {McElroy}, {Papin},
  {Plavchan}, {Ram{\'{\i}}rez}, {Rey}, {von Braun}, {Wittman}, {Abajian},
  {Ali}, {Beichman}, {Beekley}, {Berriman}, {Berukoff}, {Bryden}, {Chan},
  {Groom}, {Lau}, {Payne}, {Regelson}, {Saucedo}, {Schmitz}, {Stauffer},
  {Wyatt}, \& {Zhang}}]{Akeson13}
{Akeson}, R.~L., {Chen}, X., {Ciardi}, D., {et~al.} 2013, \pasp, 125, 989

\bibitem[{{Borucki} {et~al.}(2010){Borucki}, {Koch}, {Basri}, {Batalha},
  {Brown}, {Caldwell}, {Caldwell}, {Christensen-Dalsgaard}, {Cochran},
  {DeVore}, {Dunham}, {Dupree}, {Gautier}, {Geary}, {Gilliland}, {Gould},
  {Howell}, {Jenkins}, {Kondo}, {Latham}, {Marcy}, {Meibom}, {Kjeldsen},
  {Lissauer}, {Monet}, {Morrison}, {Sasselov}, {Tarter}, {Boss}, {Brownlee},
  {Owen}, {Buzasi}, {Charbonneau}, {Doyle}, {Fortney}, {Ford}, {Holman},
  {Seager}, {Steffen}, {Welsh}, {Rowe}, {Anderson}, {Buchhave}, {Ciardi},
  {Walkowicz}, {Sherry}, {Horch}, {Isaacson}, {Everett}, {Fischer}, {Torres},
  {Johnson}, {Endl}, {MacQueen}, {Bryson}, {Dotson}, {Haas}, {Kolodziejczak},
  {Van Cleve}, {Chandrasekaran}, {Twicken}, {Quintana}, {Clarke}, {Allen},
  {Li}, {Wu}, {Tenenbaum}, {Verner}, {Bruhweiler}, {Barnes}, \&
  {Prsa}}]{Borucki10}
{Borucki}, W.~J., {Koch}, D., {Basri}, G., {et~al.} 2010, Science, 327, 977

\bibitem[{{Campbell} {et~al.}(1988){Campbell}, {Walker}, \&
  {Yang}}]{Campbell88}
{Campbell}, B., {Walker}, G.~A.~H., \& {Yang}, S. 1988, \apj, 331, 902

\bibitem[{{Christiansen} {et~al.}(2017){Christiansen}, {Vanderburg}, {Burt},
  {Fulton}, {Batygin}, {Benneke}, {Brewer}, {Charbonneau}, {Ciardi}, {Collier
  Cameron}, {Coughlin}, {Crossfield}, {Dressing}, {Greene}, {Howard}, {Latham},
  {Molinari}, {Mortier}, {Mullally}, {Pepe}, {Rice}, {Sinukoff}, {Sozzetti},
  {Thompson}, {Udry}, {Vogt}, {Barman}, {Batalha}, {Bouchy}, {Buchhave},
  {Butler}, {Cosentino}, {Dupuy}, {Ehrenreich}, {Fiorenzano}, {Hansen},
  {Henning}, {Hirsch}, {Holden}, {Isaacson}, {Johnson}, {Knutson}, {Kosiarek},
  {L{\'o}pez-Morales}, {Lovis}, {Malavolta}, {Mayor}, {Micela}, {Motalebi},
  {Petigura}, {Phillips}, {Piotto}, {Rogers}, {Sasselov}, {Schlieder},
  {S{\'e}gransan}, {Watson}, \& {Weiss}}]{Christiansen17}
{Christiansen}, J.~L., {Vanderburg}, A., {Burt}, J., {et~al.} 2017, \aj, 154,
  122

\bibitem[{{Crossfield} {et~al.}(2017){Crossfield}, {Ciardi}, {Isaacson},
  {Howard}, {Petigura}, {Weiss}, {Fulton}, {Sinukoff}, {Schlieder}, {Mawet},
  {Ruane}, {de Pater}, {de Kleer}, {Davies}, {Christiansen}, {Dressing},
  {Hirsch}, {Benneke}, {Crepp}, {Kosiarek}, {Livingston}, {Gonzales},
  {Beichman}, \& {Knutson}}]{Crossfield17}
{Crossfield}, I.~J.~M., {Ciardi}, D.~R., {Isaacson}, H., {et~al.} 2017, \aj,
  153, 255

\bibitem[{{Danby}(1988)}]{Danby88}
{Danby}, J.~M.~A. 1988, {Fundamentals of celestial mechanics}

\bibitem[{{Deck} {et~al.}(2014){Deck}, {Agol}, {Holman}, \&
  {Nesvorn{\'y}}}]{Deck14}
{Deck}, K.~M., {Agol}, E., {Holman}, M.~J., \& {Nesvorn{\'y}}, D. 2014, \apj,
  787, 132

\bibitem[{{Eastman} {et~al.}(2013){Eastman}, {Gaudi}, \& {Agol}}]{Eastman13}
{Eastman}, J., {Gaudi}, B.~S., \& {Agol}, E. 2013, \pasp, 125, 83

\bibitem[{{Ford}(2006)}]{Ford06}
{Ford}, E.~B. 2006, \apj, 642, 505

\bibitem[{{Foreman-Mackey} {et~al.}(2013){Foreman-Mackey}, {Hogg}, {Lang}, \&
  {Goodman}}]{DFM13}
{Foreman-Mackey}, D., {Hogg}, D.~W., {Lang}, D., \& {Goodman}, J. 2013, \pasp,
  125, 306

\bibitem[{Foreman-Mackey {et~al.}(2016)Foreman-Mackey, Vousden, Price-Whelan,
  Pitkin, Zabalza, Ryan, Emily, Smith, Ashton, Cruz, Kerzendorf, Caswell,
  Hoyer, Barbary, Czekala, Rein, Gentry, Brewer, \& Hogg}]{DFM16}
Foreman-Mackey, D., Vousden, W., Price-Whelan, A., {et~al.} 2016, corner.py:
  corner.py v2.0.0, , , doi:10.5281/zenodo.53155.
\newblock \url{https://doi.org/10.5281/zenodo.53155}

\bibitem[{Fulton \& Petigura(2017)}]{radvelcite}
Fulton, B., \& Petigura, E. 2017, RadVel: Radial Velocity Fitting Toolkit, , ,
  doi:10.5281/zenodo.580821.
\newblock \url{https://doi.org/10.5281/zenodo.580821}

\bibitem[{{Fulton} {et~al.}(2016){Fulton}, {Howard}, {Weiss}, {Sinukoff},
  {Petigura}, {Isaacson}, {Hirsch}, {Marcy}, {Henry}, {Grunblatt}, {Huber},
  {von Braun}, {Boyajian}, {Kane}, {Wittrock}, {Horch}, {Ciardi}, {Howell},
  {Wright}, \& {Ford}}]{Fulton16}
{Fulton}, B.~J., {Howard}, A.~W., {Weiss}, L.~M., {et~al.} 2016, \apj, 830, 46

\bibitem[{{Gelman} {et~al.}(2003){Gelman}, {Carlin}, {Stern}, \&
  {Rubin}}]{Gelman03}
{Gelman}, A., {Carlin}, J.~B., {Stern}, H.~S., \& {Rubin}, D.~B. 2003,
  {Bayesian Data Analysis}, 2nd edn. ({Chapman and Hall})

\bibitem[{{Goodman} \& {Weare}(2010)}]{Goodman10}
{Goodman}, J., \& {Weare}, J. 2010, Communications in Applied Mathematics and
  Computational Science, Vol.~5, No.~1, p.~65-80, 2010, 5, 65

\bibitem[{{Grunblatt} {et~al.}(2017){Grunblatt}, {Huber}, {Gaidos}, {Lopez},
  {Howard}, {Isaacson}, {Sinukoff}, {Vanderburg}, {Nofi}, {Yu}, {North},
  {Chaplin}, {Foreman-Mackey}, {Petigura}, {Ansdell}, {Weiss}, {Fulton}, \&
  {Lin}}]{Grunblatt17}
{Grunblatt}, S.~K., {Huber}, D., {Gaidos}, E., {et~al.} 2017, ArXiv e-prints,
  arXiv:1706.05865

\bibitem[{{Iglesias-Marzoa} {et~al.}(2015){Iglesias-Marzoa},
  {L{\'o}pez-Morales}, \& {Jes{\'u}s Ar{\'e}valo Morales}}]{Iglesias15}
{Iglesias-Marzoa}, R., {L{\'o}pez-Morales}, M., \& {Jes{\'u}s Ar{\'e}valo
  Morales}, M. 2015, \pasp, 127, 567

\bibitem[{{Latham} {et~al.}(1989){Latham}, {Stefanik}, {Mazeh}, {Mayor}, \&
  {Burki}}]{Latham89}
{Latham}, D.~W., {Stefanik}, R.~P., {Mazeh}, T., {Mayor}, M., \& {Burki}, G.
  1989, \nat, 339, 38

\bibitem[{{Lucy} \& {Sweeney}(1971)}]{Lucy71}
{Lucy}, L.~B., \& {Sweeney}, M.~A. 1971, \aj, 76, 544

\bibitem[{{Marcy} \& {Butler}(1996)}]{Marcy96}
{Marcy}, G.~W., \& {Butler}, R.~P. 1996, \apjl, 464, L147+

\bibitem[{{Mayor} \& {Queloz}(1995)}]{Mayor95}
{Mayor}, M., \& {Queloz}, D. 1995, \nat, 378, 355

\bibitem[{{Mede} \& {Brandt}(2017)}]{Mede17}
{Mede}, K., \& {Brandt}, T.~D. 2017, \aj, 153, 135

\bibitem[{{Meschiari} \& {Laughlin}(2010)}]{Meschiari10}
{Meschiari}, S., \& {Laughlin}, G.~P. 2010, \apj, 718, 543

\bibitem[{{Meschiari} {et~al.}(2009){Meschiari}, {Wolf}, {Rivera}, {Laughlin},
  {Vogt}, \& {Butler}}]{Meschiari09}
{Meschiari}, S., {Wolf}, A.~S., {Rivera}, E., {et~al.} 2009, \pasp, 121, 1016

\bibitem[{{Murray} \& {Correia}(2010)}]{Seager10}
{Murray}, C.~D., \& {Correia}, A.~C.~M. 2010, {Keplerian Orbits and Dynamics of
  Exoplanets}, ed. S.~{Seager}, 15--23

\bibitem[{{Murray} \& {Dermott}(1999)}]{Murray99}
{Murray}, C.~D., \& {Dermott}, S.~F. 1999, {Solar system dynamics} (Solar
  system dynamics by Murray, C.~D., 1999)

\bibitem[{Newville {et~al.}(2014)Newville, Stensitzki, Allen, \&
  Ingargiola}]{Newville14}
Newville, M., Stensitzki, T., Allen, D.~B., \& Ingargiola, A. 2014, {LMFIT:
  Non-Linear Least-Square Minimization and Curve-Fitting for Python}, , ,
  doi:10.5281/zenodo.11813.
\newblock \url{https://doi.org/10.5281/zenodo.11813}

\bibitem[{{Petigura} {et~al.}(2017){Petigura}, {Sinukoff}, {Lopez},
  {Crossfield}, {Howard}, {Brewer}, {Fulton}, {Isaacson}, {Ciardi}, {Howell},
  {Everett}, {Horch}, {Hirsch}, {Weiss}, \& {Schlieder}}]{Petigura17b}
{Petigura}, E.~A., {Sinukoff}, E., {Lopez}, E.~D., {et~al.} 2017, \aj, 153, 142

\bibitem[{Powell(1964)}]{Powell64}
Powell, M. J.~D. 1964, The Computer Journal, 7, 155.
\newblock \url{+ http://dx.doi.org/10.1093/comjnl/7.2.155}

\bibitem[{{Sinukoff} {et~al.}(2017{\natexlab{a}}){Sinukoff}, {Howard},
  {Petigura}, {Fulton}, {Isaacson}, {Weiss}, {Brewer}, {Hansen}, {Hirsch},
  {Christiansen}, {Crepp}, {Crossfield}, {Schlieder}, {Ciardi}, {Beichman},
  {Knutson}, {Benneke}, {Dressing}, {Livingston}, {Deck}, {L{\'e}pine}, \&
  {Rogers}}]{Sinukoff17a}
{Sinukoff}, E., {Howard}, A.~W., {Petigura}, E.~A., {et~al.}
  2017{\natexlab{a}}, \aj, 153, 70

\bibitem[{{Sinukoff} {et~al.}(2017{\natexlab{b}}){Sinukoff}, {Howard},
  {Petigura}, {Fulton}, {Crossfield}, {Isaacson}, {Gonzales}, {Crepp},
  {Brewer}, {Hirsch}, {Weiss}, {Ciardi}, {Schlieder}, {Benneke},
  {Christiansen}, {Dressing}, {Hansen}, {Knutson}, {Kosiarek}, {Livingston},
  {Greene}, {Rogers}, \& {L{\'e}pine}}]{Sinukoff17b}
---. 2017{\natexlab{b}}, \aj, 153, 271

\bibitem[{{Teske} {et~al.}(2017){Teske}, {Wang}, {Wolfgang}, {Dai}, {Shectman},
  {Butler}, {Crane}, \& {Thompson}}]{Teske17}
{Teske}, J.~K., {Wang}, S.~X., {Wolfgang}, A., {et~al.} 2017, ArXiv e-prints,
  arXiv:1711.01359

\bibitem[{{Weiss} {et~al.}(2017){Weiss}, {Deck}, {Sinukoff}, {Petigura},
  {Agol}, {Lee}, {Becker}, {Howard}, {Isaacson}, {Crossfield}, {Fulton},
  {Hirsch}, \& {Benneke}}]{Weiss17}
{Weiss}, L.~M., {Deck}, K.~M., {Sinukoff}, E., {et~al.} 2017, \aj, 153, 265

\bibitem[{{Wright} \& {Howard}(2009)}]{WrightHoward09}
{Wright}, J.~T., \& {Howard}, A.~W. 2009, \apjs, 182, 205

\end{thebibliography}

\end{document}